\documentstyle[aps]{revtex}

\begin{document}
\twocolumn[\hsize\textwidth\columnwidth\hsize\csname
@twocolumnfalse\endcsname

\title{Magnetic tension and the geometry of the universe}

\author{Christos G. Tsagas}
\address{Relativity and Cosmology Group, Division of Mathematics
and Statistics,\\ Portsmouth University, Portsmouth~PO1~2EG,
England}

\maketitle

\vspace{3mm}

\begin{abstract}
The vector nature of magnetic fields and the geometrical
interpretation of gravity introduced by general relativity,
guarantee a special coupling between magnetism and spacetime
curvature. This magneto-geometrical interaction effectively
transfers the tension properties of the field into the spacetime
fabric, triggering a variety of effects with profound
implications. Given the ubiquity of magnetic fields in the
universe, these effects could prove critical. We discuss the
nature of the magneto-curvature coupling and illustrate some of
its potential implications for cosmology.\\\\ PACS number(s):
98.80.Hw, 04.40.Nr, 95.30.Qd, 98.62.En
\end{abstract}\vskip2pc]

Despite the widespread presence of magnetic fields in the universe
\cite{K}, studies of their potential cosmological implications
remain relatively underdeveloped. It has long been thought,
however, that magnetic fields might have played a role during the
formation and the evolution of the observed large scale structure
\cite{RR}. Recently, this idea has received renewed interest
manifested by the increasing number of related papers that have
appeared in the literature \cite{KOR,TB}. Nevertheless, there are
still only a few fully relativistic approaches available. Most
treatments are either Newtonian or semi-relativistic. As such,
they are bound to exclude certain features of the magnetic nature.
Two key features are the vector nature of the field and the
tension properties of magnetic force lines. In general relativity
vector fields have quite a different status than ordinary scalar
sources, such as the energy density and pressure of matter. The
geometrical nature of Einstein's theory guarantees that vectors
are directly coupled to the spacetime curvature. This special
interaction is manifested in the Ricci identity
\begin{equation}
2\nabla_{[a}\nabla_{b]}B_c=R_{abcd}B^d\,,  \label{Ricci}
\end{equation}
applied here to the magnetic vector $B_a$, where $R_{abcd}$ is the
spacetime Riemann tensor. The Ricci identity plays a fundamental
role in the mathematical formulation of general relativity.
Essentially, it is the definition of spacetime curvature itself.
The Ricci identity also leads to a direct coupling between
magnetism and spatial geometry. Indeed, projecting
Eq.~(\ref{Ricci}) into the instantaneous rest space of a comoving
observer we arrive at
\begin{equation}
2{\rm D}_{[a}{\rm D}_{b]}B_c= -2\omega_{ab}h_c{}^d\dot{B}_d+ {\cal
R}_{dcba}B^d\,. \label{3Ricci}
\end{equation}
In the above ${\rm D}_a$ is the projected covariant derivative
operator, $\omega_{ab}$ is the vorticity tensor and $h_{ab}$,
${\cal R}_{abcd}$ are respectively the metric and the Riemann
tensor of the observer's rest space. Note that the overdot
indicates differentiation along the observer's worldline. The
vorticity term appears because generally the observer's motion is
not hypersurface orthogonal. For our purposes, however, the key
quantity is the last one on the right-hand side of
Eq.~(\ref{3Ricci}). Its presence illustrates the direct coupling
between magnetic fields and spatial geometry. We will call this
special interaction the {\em magneto-curvature coupling}. This
coupling goes beyond the standard interplay between matter and
geometry as introduced by the Einstein field equations. In fact,
it makes the magnetic field an inseparable part of the spacetime
fabric by effectively transferring its properties to the spacetime
itself. The key property appears to be the tension of the magnetic
lines of force.

Magnetic fields transmit stresses between regions of material
particles and fluids. The field exerts an isotropic pressure in
all directions and carries a tension along the magnetic lines of
force. Each small flux-tube behaves like an infinitely elastic
rubber band, while neighbouring tubes expand against each other
under their own pressure. Equilibrium exists only when a balance
between pressure and tension is possible. To unravel these tension
properties, let us consider a pure magnetic field. Its
energy-momentum tensor decomposes as
\begin{equation}
T_{ab}={\textstyle{1\over2}}B^2u_au_b+
{\textstyle{1\over6}}B^2h_{ab}+ \Pi_{ab}\,,  \label{mTab}
\end{equation}
where $B^2=B_aB^a$ and $\Pi_{ab}=(B^2/3)h_{ab}-B_aB_b$. Thus, the
field behaves as an imperfect fluid with energy density $\rho_{\rm
m}=B^2/3$, isotropic pressure $p_{\rm m}=B^2/6$ and anisotropic
pressure $\Pi_{ab}$. Note that, in the absence of electric fields,
the electromagnetic Poynting vector is zero and Eq.~(\ref{mTab})
contains no energy-flux vector. The tension properties of the
field are incorporated in the symmetric trace-free tensor
$\Pi_{ab}$. They emerge when we take the eigenvalues of $\Pi_{ab}$
orthogonal and along the direction of the magnetic force lines.
Orthogonal to $B_a$ one finds two positive eigenvalues equal $1/3$
each. Thus, the magnetic pressure perpendicular to the field lines
is positive, reflecting their tendency to push each other apart.
In the $B_a$ direction, however, the associated eigenvalue is
$-2/3$ and the magnetic pressure is negative. The minus sign
reflects the tension properties of the field lines and their
tendency to remain as `straight' as possible.

The magnetic effects on the fluid propagate through Euler's
formula. For a barotropic, infinitely conductive, magnetised
medium the non-linear Euler equation is \cite{TB}
\begin{eqnarray}
\left(\rho+p+{\textstyle{2\over3}}B^2\right)A_a&=&-c_{\rm s}^2{\rm
D}_a\rho- \varepsilon_{abc}B^b{\rm
curl}B^c\nonumber\\&{}&-A^b\Pi_{ba}\,, \label{Euler}
\end{eqnarray}
where $A_a$ is the fluid 4-acceleration, $c_{\rm
s}^2=\dot{p}/\dot{\rho}$ is the sound speed squared and
$\varepsilon_{abc}$ the spatial alternating tensor. The projected
gradient of Eq.~(\ref{Euler}) facilitates a more detailed study of
the magnetic behaviour. In particular, it separates the effects of
the field's tension from those of the ordinary (i.e.~the positive)
magnetic pressure. To illustrate how and also to reveal the role
of the spacetime curvature, we consider a weakly magnetised
Bianchi~I background. Its generic anisotropy will help us to
identify the magnetic tension contributions easier. Employing
Eq.~(\ref{3Ricci}) we arrive at the linear expression \cite{TM}
\begin{eqnarray}
\rho(1+w){\rm D}_bA_a&=&-c_{\rm s}^2{\rm D}_b{\rm D}_a\rho-
\frac{c_{\rm s}^2}{\rho(1+w)}B_aB^c{\rm D}_b{\rm
D}_c\rho\nonumber\\&{}& +B^c{\rm D}_c{\rm D}_bB_a
-{\textstyle{1\over2}}{\rm D}_b{\rm D}_aB^2\nonumber\\&{}& -
{\textstyle{4\over3}}\Theta B_aB^c\omega_{bc} +{\cal
R}_{acbd}B^cB^d\,, \label{DbAa}
\end{eqnarray}
with $w=p/\rho$. Let us concentrate on the four magnetic terms at
the end of Eq.~(\ref{DbAa}). The first two also appear in
Newtonian studies, whereas the last two are the relativistic
corrections. Since ${\rm D}^aB_a=0=B^aB^b\omega_{ab}$, the first
standard term and the first of the relativistic corrections are
irrelevant for our purposes. Here we focus upon the
magneto-curvature term at the end of Eq.~(\ref{DbAa}), where the
spatial curvature tensor is twice contracted along the field
lines. The directional dependence on $B_a$ ensures that ${\cal
R}_{acbd}B^cB^d$ also conveys the magnetic tension effects. This
is also implied by the sign difference between the
magneto-geometrical term and the standard gradient ${\rm D}_b{\rm
D}_aB^2$, which carries the effects of the ordinary magnetic
pressure. Note that the magneto-curvature stress in
Eq.~(\ref{DbAa}) is always normal to the field lines, as the
symmetries of the Riemann tensor confirm. Also, it closely
resembles the classical curvature stress exerted by distorted
magnetic field lines (see e.g.~\cite{P}). This resemblance becomes
more apparent when we linearise (\ref{DbAa}) about a FRW
background (see Eqs.~(\ref{Re}), (\ref{Re1})). The difference is
that, in the relativistic case, the distortion of the field
pattern is triggered by the spacetime geometry itself. In fact,
the magneto-curvature term in Eq.~(\ref{DbAa}) effectively injects
the tension properties of the field into the spacetime fabric. The
implications are widespread and far from trivial.

Consider a general spacetime filled with a magnetised, highly
conductive, perfect fluid. Its volume expansion is governed by the
non-linear Raychaudhuri equation \cite{TB}
\begin{eqnarray}
{\textstyle{1\over3}}\Theta^2q&=&
{\textstyle{1\over2}}\left(\rho+3p+B^2\right)+
2\left(\sigma^2-\omega^2\right)\nonumber\\&{}&-\nabla^aA_a-
\Lambda\,, \label{Ray}
\end{eqnarray}
where $q$ is the deceleration parameter, $\sigma^2$ and $\omega^2$
are the shear and vorticity magnitudes respectively and $\Lambda$
is the cosmological constant. The state of the expansion is
determined by the sign of the right-hand side of Eq.~(\ref{Ray}).
Positive terms decelerate the universe while negative ones lead to
acceleration. Clearly, conventional matter and shear effects slow
the expansion down. On the other hand, vorticity and a positive
cosmological constant accelerate the universe. Hence, every term
on the right hand side of Eq.~(\ref{Ray}) has a clear kinematical
role with the exception of $\nabla^aA_a$. The latter can be either
positive or negative, depending on the specific form of the
4-acceleration. In our case $A_a$ obeys the non-linear Euler
formula given by Eq.~(\ref{Euler}). In a weakly magnetised,
slightly inhomogeneous and anisotropic, almost-FRW universe
Eqs.~(\ref{Euler}) and (\ref{Ray}) linearise to give \cite{MT}
\begin{eqnarray}
{\textstyle{1\over3}}\Theta^2{\rm q}&=&
{\textstyle{1\over2}}\rho(1+3w)+ \frac{c_{\rm
s}^2\Delta}{(1+w)a^2}+ \frac{c_{\rm a}^2{\cal
B}}{2(1+w)a^2}\nonumber\\&{}&-\frac{2kc_{\rm a}^2}{(1+w)a^2}-
\Lambda\, \label{Re}
\end{eqnarray}
on using the trace of (\ref{DbAa}). In the above $\Delta$ and
${\cal B}$ describe scalar perturbations in the fluid and the
magnetic energy densities respectively, $c_{\rm a}^2=B^2/\rho$ is
the square of the Alfv\'en speed, $k=0\,,\pm1$ is the background
curvature index and $a$ is the scale factor. Given that in the
linear regime the mean values of $\Delta$ and ${\cal B}$ are zero,
one expects that on average Eq.~(\ref{Re}) looks like
\begin{equation}
{\textstyle{1\over3}}\Theta^2{\rm q}=
{\textstyle{1\over2}}\rho(1+3w)- \frac{2kc_{\rm a}^2}{(1+w)a^2}\,,
\label{Re1}
\end{equation}
where $\Lambda=0$ from now on. Note the magneto-curvature term in
the right-hand side which results from the coupling between
magnetism and geometry as manifested in Eq.~(\ref{3Ricci}). This
term affects the expansion in two completely different ways
depending on the sign of the background curvature. In particular,
the magneto-geometrical effects slow the expansion down when
$k=-1$ but tend to accelerate the expansion if $k=+1$. Such a
behaviour seems odd, especially since positive curvature is always
associated with gravitational collapse. The explanation lies in
the elastic properties of the field lines. As curvature distorts
the magnetic force lines their tension backreacts giving rise to a
restoring magneto-curvature stress \cite{TM}. The magnetic
backreaction has kinematical, dynamical as well as geometrical
implications. In Eq.~(\ref{Re1}), for example, the tension of the
field adjusts the expansion rate of the universe to minimize the
kinematical effects of curvature. As a result the expansion rate
is brought closer to that of a flat FRW model. Overall, it looks
as though the elastic properties of the field have been
transferred into space. According to Eq.~(\ref{Re1}), the
magneto-curvature effects also depend on the material component of
the universe. When dealing with conventional matter (i.e.~for
$0\leq w\leq1$) the most intriguing cases occur in positively
curved spaces \cite{MT}. In particular, when $w=1$ (i.e.~for stiff
matter) the Alfv\'en speed grows as $c_{\rm a}^2\propto a^2$ and
the magneto-curvature term in Eq.~(\ref{Re1}) becomes
time-independent. In this case the field acts as an effective
positive cosmological constant. For radiation and dust, on the
other hand, $c_{\rm a}^2={\rm const.}$ and $c_{\rm a}^2\propto
a^{-1}$ respectively. In these cases the magneto-curvature term is
no longer time independent but drops with time mimicking a
time-decaying quintessence. The coupling between magnetism and
geometry also means that even weak magnetic fields can have a
strong impact if the curvature is strong. To demonstrate how this
might happen, consider a weakly magnetised spatially open
cosmology filled with non-conventional matter (i.e.~$k=-1$ and
$-1\leq w<0$). Scalar fields, for example, can have an effective
equation of state that satisfies this requirement. Such models
allow for an early curvature dominated regime with $\Omega\ll1$.
Given that $\rho\propto a^{-3(1+w)}$ and $c_{\rm a}^2\propto
a^{-1+3w}$, the magneto-curvature term in Eq.~(\ref{Re1}) can
dominate the early expansion, even when the field is weak, if
$-1\leq w\leq-1/3$. In this case the accelerated inflationary
phase, which otherwise would have been inevitable, is suppressed.
Instead of inflating the magnetised universe remains in a state of
decelerated expansion. For $w=-1$, in particular, the mere
presence of the field can inhibit the de Sitter inflationary
regime if $\Omega<0.5$ \cite{MT}. This result has two
implications. First, it challenges the widespread perception that
magnetic fields are relatively unimportant for cosmology. Even
weak fields can play a decisive role when the curvature is strong.
Second, it casts doubt on the efficiency of standard inflation in
the presence of primordial magnetism.

Let us now turn our attention to geometry and examine the
implications of the magnetic tension for propagating gravitational
radiation. To begin with, recall the tendency of the field lines
to remain straight. If this property were transferred to the
spacetime, we would expect to see a suppressing effect on gravity
waves propagating through a magnetised region. Such damping should
appear as a decrease in the wave's energy density and amplitude.
To put this idea to the test we consider linear gravitational
waves in a weakly magnetised almost-FRW universe, filled with a
highly conductive medium. We also assume that the background
spatial sections are flat and we address superhorizon scales only.
Covariantly, gravity waves are described via the electric
($E_{ab}$) and the magnetic ($H_{ab}$) parts of the Weyl tensor
\cite{E}. Their magnitudes, $E^2=E_{ab}E^{ab}/2$ and
$H^2=H_{ab}H^{ab}/2$, provide a measure of the wave's energy
density and amplitude. Given that $H_{ab}={\rm curl}\sigma_{ab}$,
we can simplify the problem by replacing the magnetic Weyl tensor
with the shear. Note that the field couples to gravitational
radiation directly via the anisotropic magnetic stresses, which
affect the propagation of both $E_{ab}$ and $\sigma_{ab}$
\cite{MTU}. Having set the constraints that isolate tensor
perturbations in a magnetised universe (see \cite{MTU}), we arrive
at the system
\begin{eqnarray}
(E^2)^{.}&=&-2\Theta E^2- {\textstyle{1\over2}}\rho(1+w){\cal X}-
{\textstyle{1\over2}}\Theta B^2{\cal E}\,,\nonumber\\
(\sigma^2)^{.}&=&-{\textstyle{4\over3}}\Theta\sigma^2- {\cal X}
-{\textstyle{1\over2}}B^2\Sigma\,,\nonumber\\ \dot{{\cal X}}&=&-
{\textstyle{5\over3}}\Theta{\cal X}- 2E^2-
\rho(1+w)\sigma^2\nonumber\\&{}&- {\textstyle{1\over2}}B^2{\cal
E}-{\textstyle{1\over2}}\Theta B^2\Sigma\,,\nonumber\\ \dot{{\cal
E}}&=&-\Theta{\cal E}- {\textstyle{1\over2}}\rho(1+w)\Sigma-
{\textstyle{1\over3}}\Theta B^2\,,\nonumber\\
\dot{\Sigma}&=&-{\textstyle{2\over3}}\Theta\Sigma-{\cal
E}-{\textstyle{1\over3}}B^2\,,  \label{dotSigma}
\end{eqnarray}
with $B^2\propto a^{-4}$, ${\cal X}=E_{ab}\sigma^{ab}$, ${\cal
E}=E_{ab}\eta^a\eta^b$ and $\Sigma=\sigma_{ab}\eta^a\eta^b$
($\eta_a=B_a/\sqrt{B^2}$). The last two scalars are related via
the Gauss-Codacci equation by
\begin{equation}
{\cal E}={\textstyle{1\over3}}\Theta\Sigma+
{\textstyle{1\over3}}B^2+ {\rm R}\,,  \label{GC}
\end{equation}
where ${\rm R}=[{\cal R}_{(ab)}-({\cal R}/3)h_{ab}]\eta^a\eta^b$
describes spatial curvature distortions in the direction of the
magnetic field lines. For radiation, the late-time solution for
$E^2$ is \cite{MTU}
\begin{eqnarray}
E^2&=&{\textstyle{4\over9}}
\left[E_0^2+\frac{\sigma_0^2}{4t_0^2}-\frac{{\cal
X}_0}{2t_0}\right] \left(\frac{t_0}{t}\right)^2\nonumber\\&{}&-
{\textstyle{2\over9}}\left({\textstyle{1\over6}}B_0^2+{\rm R}_0
\right)B_0^2 \left(\frac{t_0}{t}\right)^2\,, \label{E2}
\end{eqnarray}
with an analogous result for dust \cite{MTU}. Note that the term
in square brackets determines the magnetic-free case. According to
Eq.~(\ref{E2}), the field leaves the evolution rate of $E^2$
unchanged but modifies its magnitude. The magnetic impact is
twofold. There is a pure magnetic effect, independent of the
spatial curvature, which always suppresses the energy of the wave.
It becomes apparent when we set ${\rm R}_0=0$ in Eq.~(\ref{E2}).
This effect is the direct result of the magnetic tension. As the
wave propagates it distorts the field lines which backreact by
smoothing out any ripples in the spacetime fabric. The
magnetically induced damping is proportional to the ratio
$B^2_0/E_0$. Given the inherent weakness of gravitational
radiation, the magnetic effects are potentially detectable even
when relatively weak fields are involved. Solution (\ref{E2}) also
reveals a magneto-curvature effect on gravitational radiation.
This is encoded in the ${\rm R}_0$-term and depends entirely on
the spatial curvature. For ${\rm R}_0>0$, namely when the
curvature distortion along the field lines is positive, the
pure-magnetic damping is further enhanced. On the other hand, the
suppressing effect of the field weakens if ${\rm R}_0<0$. In fact,
the field will increase the energy of the wave provided that ${\rm
R}_0<-B_0^2/6$. These magneto-geometrical effects get stronger
with increasing curvature distortion. Let us take a closer look at
them. According to Eq.~(\ref{GC}), the scalar ${\rm R}$ describes
distortions in the local spatial curvature generated by the
propagating magnetised gravity wave. Clearly, the
magneto-geometrical term in Eq.~(\ref{E2}) modifies the energy
density of the wave in a way that always minimizes such curvature
distortions. In other words, the pure magneto-curvature effect
tends to preserve the spatial flatness of the background universe.
Earlier, an analogous magneto-curvature effect was also observed
on the expansion rate of spatially curved FRW universes. This
pattern of behaviour raises the question as to whether it reflects
a generic feature of the magnetic nature. More specifically, one
wonders if the tension properties of the magnetic force lines and
the coupling between magnetism and spacetime curvature, imply an
inherent `preference' of the field for flat geometry. Let us take
a more direct look at this possibility. Consider an almost-FRW
magnetised universe and assume that the background spatial
geometry is Euclidean. If ${\cal R}$ is the Ricci scalar of the
perturbed spatial sections, then using the trace of
Eq.~(\ref{DbAa}) we obtain \cite{TB}
\begin{eqnarray}
\dot{{\cal R}}&=&-{\textstyle{2\over3}}\left[1+\frac{2c_{\rm
a}^2}{3(1+w)}\right]\Theta{\cal R}+ \frac{4c_{\rm
s}^2\Theta}{3(1+w)a^2}\Delta\nonumber\\&{}&+\frac{2c_{\rm
a}^2\Theta}{3(1+w)a^2}{\cal B}\,.  \label{dotcR}
\end{eqnarray}
As expected, the expansion dilutes curvature distortions, caused
by flactuations in the fluid and the magnetic energy densities.
Interestingly, the field also enhances the smoothing effect of the
expansion. This effect results from the tension properties of the
magnetic force lines, which tend to suppress curvature
distortions. Given the weakness of the field (recall that $c_{\rm
a}^2\ll1$), the magnetically induced smoothing is negligible
compared to that caused directly by the expansion. Nevertheless,
the tendency of the field to maintain the original flatness of the
spatial sections is quite intriguing. It seems to support the idea
that, given their tension properties and their direct coupling to
curvature, magnetic fields might indeed have a natural preference
for flat spaces.

The magneto-curvature effects presented here reveal a side of the
magnetic nature which as yet remains unexplored. They derive from
the vector nature of the field and from the geometrical approach
to gravity adopted by general relativity. The latter allows a
direct coupling between magnetism and curvature which effectively
transfers the magnetic properties into space itself. The tension
of the field lines appears to be the key property. Kinematically
speaking, the magneto-curvature effects tend to accelerate
spatially closed regions, while they decelerate those with open
spatial curvature. Crucially, if the curvature input is strong,
the overall impact can also be strong even when the field is weak.
This challenges the widespread belief that, due to their perceived
weakness, magnetic fields are relatively unimportant for
cosmology. Inflationary scenarios allow for a strong-curvature
regime during their very early stages. An initial curvature
dominated epoch has never been considered a serious problem for
inflation given the vast smoothing power of the accelerated
expansion. It is during these early stages, however, that a weak
magnetic presence is found capable of suppressing the accelerated
phase in spatially open `inflationary' models. Such a result casts
doubt on the efficiency, and potentially on the viability, of
standard inflation in the presence of primeval magnetism. In fact,
every cosmological model that allows for a strong-curvature regime
and a weak magnetic field could be vulnerable to these
magneto-curvature effects. The coupling between magnetism and
spacetime curvature has also intriguing geometrical implications.
It modifies the expansion of spatially closed, and open, FRW
universes bringing the rate closer to that of a flat Friedmannian
model. The tension of the field lines is found to suppress
gravitational waves propagating through a magnetised region.
Moreover, the combined magneto-curvature effects smooth out
perturbations in the spatial curvature of a flat FRW universe, and
modulate the energy of gravity waves as if to preserve the
background flatness. In short the magnetised space seems to react
to curvature distortions showing, what one might interpret as, a
preference for flat geometry. Given the ubiquity of magnetic
fields in the universe, this unconventional behaviour deserves
further investigation as it could reflect a deeper
inter-connection between electromagnetism and geometry. This in
turn could drastically change our views on the role of cosmic
magnetism not only only in cosmology but also in astrophysics. It
is the aim of this letter to bring these issues to light and draw
attention to their potential implications.\\

The author wishes to thank A.A. Coley, R. Maartens, D.R. Matravers
and L. Mestel for helpful discussions and comments. This work was
supported by PPARC.

\end{document}